\documentclass[a4paper,12pt,final]{article}

\usepackage[sort&compress]{natbib}
\setcitestyle{open={},close={},super}
\usepackage{hyperref}
\usepackage[T1]{fontenc}       
\usepackage{authblk}
\usepackage{amsmath}
\usepackage{amsfonts}
\usepackage{amssymb}
\usepackage{graphicx}
\usepackage{hyperref}
\usepackage{xcolor}
\usepackage{ulem}

\author[1]{M. G. Damaceno\footnote{damascenomaria863@gmail.com}}
\author[2]{A. Mucherino\footnote{antonio.mucherino@irisa.fr}}
\author[1]{R. Medeiros de Ara\'{u}jo\footnote{rennene@gmail.com}}
\author[1]{P. H. Souto Ribeiro\footnote{p.h.s.ribeiro@ufsc.br}}
\author[1]{N. Rubiano da Silva\footnote{nara.rubiano@ufsc.br}}
\affil[1]{Departamento de F\'{i}sica, Universidade Federal de Santa Catarina, CEP 88040-900, Florian\'{o}plis, SC, Brazil}
\affil[2]{IRISA, University of Rennes 1, Rennes, France}

\date{}

\title{Experimental Investigation of Optical Processing With Spatial Light Modulation}

\begin{document}
	
\maketitle


 \begin{abstract}
The growing demand for real-time data processing in applications such as neural networks and embedded control systems has spurred the search for faster, more efficient alternatives to traditional electronic systems. In response, we experimentally investigate an optical processing scheme that encodes information in the transverse wavefront of light fields using spatial light modulators. Our goal is to explore the limits on parallelism imposed by technical constraints. We begin by implementing optical an XOR logic gate applied to binary matrices. By analyzing the average error rate in the optical operation between matrices of varying sizes (up to 300 × 300 elements), we analyze the bit depth capacity of the system and the role of information redundancy. Furthermore, we successfully demonstrate image encryption and decryption using a one-time pad protocol for matrices as large as 164 × 164 elements. These findings support the development of a high-dimensional matrix optical processor.

\end{abstract}


\section{Introduction}

Electronic information processing has driven significant advances in computing capabilities, as exemplified by Moore's law \cite{moore1975progress,moore1998cramming}. However, it faces thermodynamic limitations due to miniaturization and heat dissipation \cite{packan1999pushing,pop2006heat,yan1992scaling,markov2014limits,cao2023future}. Exploring alternative processor architectures is a promising approach to address these challenges, with optical platforms emerging as candidates for efficient and high-speed computing \cite{bell1986optical,silva2014performing,McMahon2023}. These platforms have shown potential for both general computation and neural network implementations \cite{lin2018all,colburn2019optical,hamerly2019large,feldmann2021parallel,zuo2021scalability}, as well as for emulating quantum computation \cite{koni2024emulating}.

Recent advances include on-chip tensor-core-based \cite{feldmann2021parallel} and interference-based \cite{shen2017deep} optical systems, in addition to free-space diffraction-based neural networks demonstrating high parallel computing capability and power efficiency both in all-optical \cite{lin2018all} as in hybrid photonic-electronic \cite{colburn2019optical} architectures. A particularly flexible approach leverages the transverse spatial degrees of freedom of light beams for optical parallelism \cite{Hengeveld2022,zuo2021scalability}. In such schemes, the degree of parallelism is fundamentally limited by the spatial resolution achievable in phase and amplitude modulation. When using spatial light modulators (SLMs), the constraints are determined by the number and size of pixels that can be simultaneously addressed by a single optical wavefront.

Here, we present an experimental investigation and quantitative evaluation of the fundamental limitations in a free-space optical system that encodes information in the transverse spatial degrees of freedom of light, following the architecture proposed in \cite{Hengeveld2022}. We employ polarization as an ancillary degree of freedom to convert phase modulation into amplitude modulation \cite{davis2000twodimensional}.

Specifically, we assess the system's performance for optical encryption based on XOR operations within the one-time pad cryptographic protocol. These XOR operations are implemented optically using cascaded spatial light modulators (SLMs) \cite{han1999optical, unnikrishnan2000polarization, mogensen2000phase, tu2004optical, kumar2019image}. By systematically varying the number of SLM pixels used to represent each binary variable, we examine the practical limits of parallelization in optical computing and evaluate their impact on encryption performance.

Our results reveal that the scalability of optical XOR-based encryption is fundamentally limited by the pixel density requirements representing a variable, with significant signal degradation occurring beyond a critical threshold. These findings provide crucial insights into the parallelization constraints of this computational approach and elucidate the trade-offs between scalability, performance, and system complexity. Furthermore, they indicate strategies for overcoming these limitations and achieving higher parallelism levels, ultimately approaching the technical limit of one variable per SLM pixel.


\section{Experimental setup}

\begin{figure}
    \centering
    \includegraphics[width=.7\linewidth]{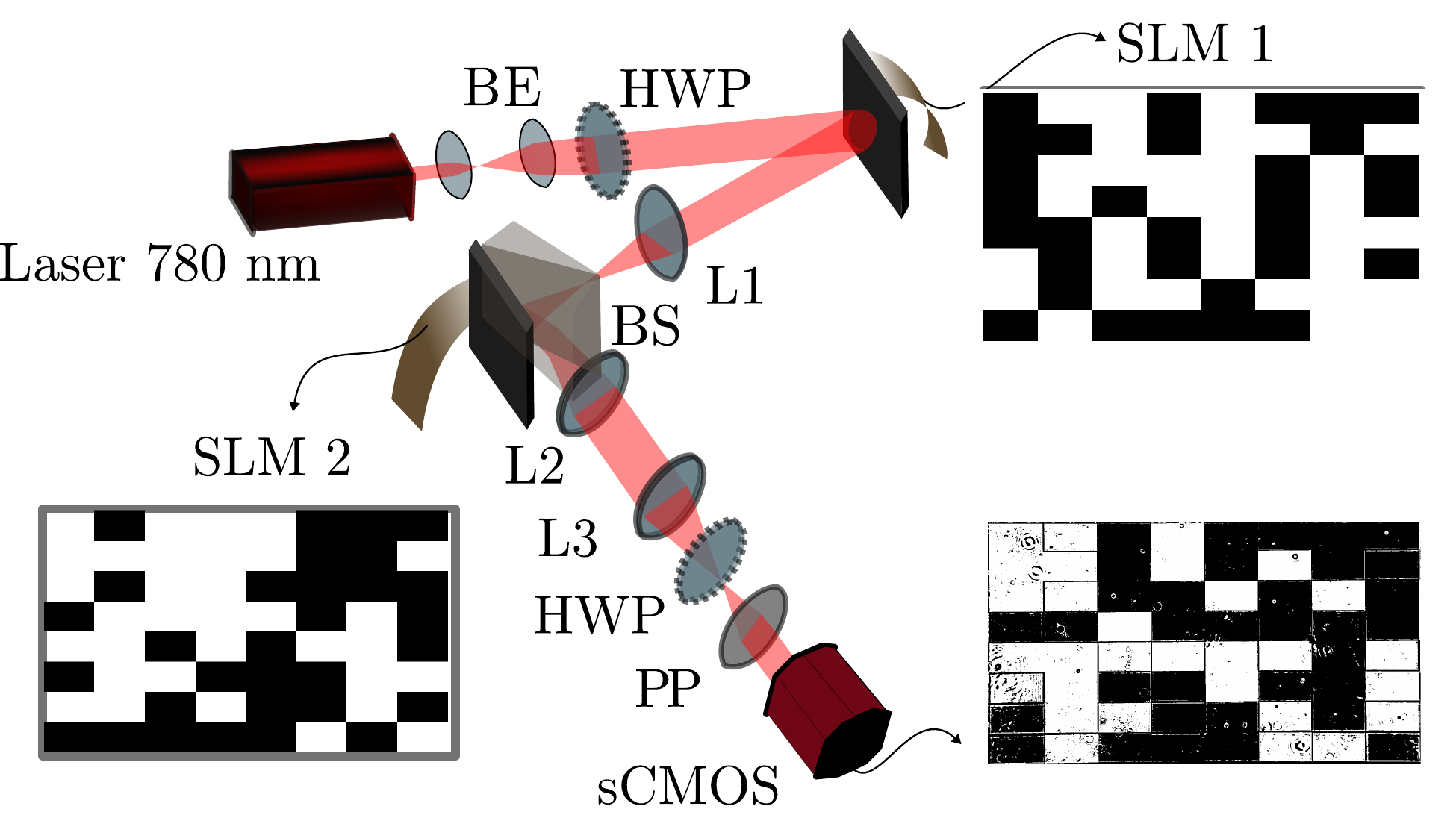}
    \caption{Scheme of the experimental setup: BE (beam expander); SLM1 and SLM2 (spatial light modulators); L1, L2 and L3 (lenses); BS (beam splitter); HWP (half-wave plate); PP (polymer polarizer); and sCMOS camera. The image projected on both SLMs represents binary information encoded in blocks of pixels (insets).}
    \label{fig:sch-setup}
\end{figure}

The experimental setup is sketched in Figure~\ref{fig:sch-setup}. The light source is a laser diode oscillating at 780 nm, which is expanded and collimated. It is directed to a spatial light modulator (SLM) and its diameter is prepared in order to fully illuminate the SLM surface. With that, we ensure a nearly flat wavefront and nearly homogeneous intensity impinging on the SLM. We use an SLM Holoeye Gaea-2, with resolution 3840$\times$2160 pixels and active area with a diagonal of $\sim$ 1.8 cm. Let us call it SLM1. 

The screen of SLM1 is imaged onto a second SLM, called SLM2. It is a Holoeye Leto-3, with 1920$\times$1080 resolution. The imaging system is such that the image of the screen of SLM1 completely matches the screen of SLM2. As a result, there is a pixel mapping ratio of 4:1 from SLM1 to SLM2, and the maximal resolution of the images used for testing the setup is full HD (1920$\times$1080), which corresponds to the resolution of SLM2. A 2-inch-diameter lens (L1) with focal distance 200 mm is used in the imaging scheme instead of the more usual 1-inch-diameter, in order to minimize diffraction effects and to improve light collection from the relatively large SLM area. The screen of SLM2 is imaged onto a sCMOS camera with a pixel pitch of 5.04 µm and full HD resolution. Before the light from SLM2 reaches the camera it goes through a half-wave plate (HWP) and a polymer polarizer (PP).

To implement a XOR operation with this setup, we use the incident light’s polarization to convert the phase patterns from the SLMs into amplitude variations, allowing detection with a simple camera, as explained below. 

We employ incident light with a diagonal polarization at \(45^\circ\), described by the Jones vector:
\begin{equation}
\mathbf{E}_{\text{in}} = \frac{1}{\sqrt{2}} \begin{bmatrix} 1 \\ 1 \end{bmatrix}.
\end{equation}
The light is sequentially reflected by SLM1 and SLM2, which only modulate the horizontal polarization component. Each SLM is set to apply a phase shift of either \(0\) or \(\pi\) with respect to the vertical component, pixel by pixel. The phase of each pixel can be independently set, and is indexed by $l,m$. This binary modulation corresponds to the logical states \(0\) and \(1\) (black and white, respectively). The following Jones matrix describes the combined effect of the two SLMs:

\begin{equation}\label{Jslm}
\mathbf{J}_{\text{SLMs}} = \begin{bmatrix} 
    \exp\left(j(\phi_{lm} + \phi'_{lm})\right) & 0 \\ 
    0 & 1 
\end{bmatrix},
\end{equation}
where \(\phi_{lm}\) and \(\phi'_{lm}\) are the phase shifts applied by the first and second SLM, respectively.

After modulation, the light encounters a half-wave plate (HWP) followed by a polymer polarizer (PP). The HWP rotates the polarization states, converting diagonal (D) to horizontal (H) polarization and antidiagonal (A) to vertical (V) polarization. The PP then filters the polarization, allowing only the vertical component to reach the camera. This optical setup effectively maps the polarization states to binary intensity levels detected by the camera.

Since the Jones matrix for the HWP and PP is:
\begin{equation}\label{eq:Jpolfilt}
\mathbf{J}_{\text{pol}} = 
\begin{bmatrix} 
    0 & 0 \\ 
    0 & 1 
\end{bmatrix}
\dfrac{1}{\sqrt{2}}\begin{bmatrix} 
    1 & 1 \\ 
    1 & -1 
\end{bmatrix},
\end{equation}
the electric field after the PP is given by:
\begin{equation}
\mathbf{E}^{lm}_{\text{out}} = \mathbf{J}_{\text{Total}} \cdot \mathbf{E}_{\text{in}} = \frac{1}{2} \begin{bmatrix} 
    0\\ 
    \exp\left(j(\phi_{lm} + \phi'_{lm})\right) - 1  
\end{bmatrix}.
\label{eq:Eout}
\end{equation}

We use a sCMOS camera to measure the intensity distribution after polarization projection and filtering. The recorded intensity, $I_V$, is proportional to $|\exp\left(j(\phi_{lm} + \phi'_{lm})\right) - 1|^2$. Therefore, $I_V = 0$ (black) for $\phi_{lm} + \phi'_{lm}= 0$ or $2\pi$ and  $I_V = \text{maximum}$ (white) for $\phi_{lm} + \phi'_{lm}= \pi$. In this way, the logical XOR operation is realized.

The setup performance for XOR operations was evaluated by encoding binary matrices into polarization states and comparing the recorded images with theoretical results. The system encoding capacity was tested for matrix sizes ranging from $10 \times 10$ to $300 \times 300$. Each $N\times N$ matrix was displayed by splitting the whole SLM screen evenly into $N\times N$ rectangles \cite{supplement}.


\section{Results}

\subsection{Optical resolution}
The resolution of our optical system is limited by diffraction and aberration. Considering the wavelength $\lambda$ of 780~nm 
and that the lens between the SLMs has a focal length $f=200$~mm and a diameter $D=50.8$~mm ($2^{''}$), the minimum resolvable separation is $\Delta l = 1.22 \frac{\lambda f}{D} = 3.74$~$\mu$m (center-to-center). This provides an ideal lower bound that allows addressing individual SLM pixels, even for the 4K Gaea Holoeye SLM, which has pixel size of 6.4~$\mu$m.

\begin{figure}
    \centering
    \includegraphics[width=.7\linewidth]{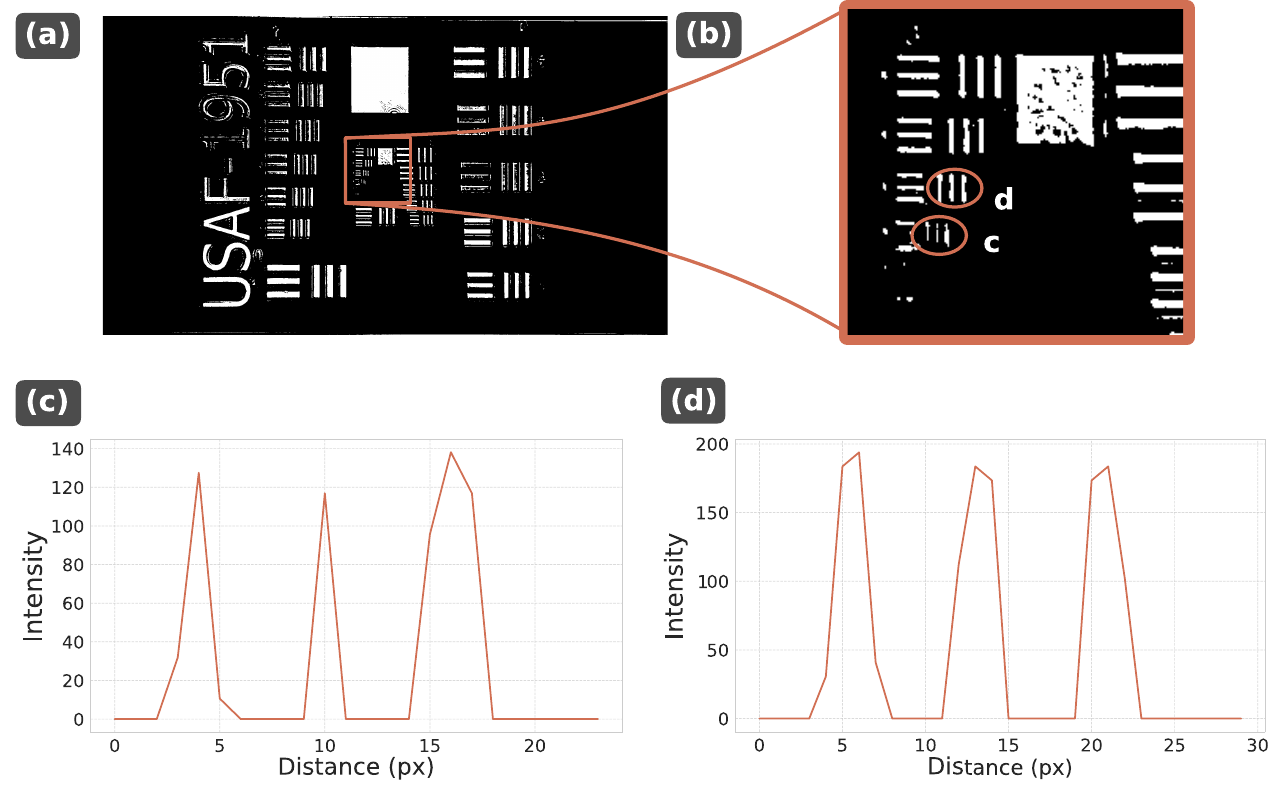}
    \caption{Recorded image of the USAF-1951 test chart displayed on the first SLM for verifying optical resolution. (a) Full image, and (b) a zoom of the smallest pixels blocks. Integrated intensity profiles of the (c) 3-pixel and (d) 4-pixel spacing blocks.}
    \label{fig:usaf}
\end{figure}

We test the actual resolution of our setup by projecting the full HD of an adapted USAF-1951 test chart, consisting of horizontal and vertical $n$-pixel bars, onto the first, 4K SLM (SLM1). The reflected light is sent to SLM2, on which a flat mask is displayed, and then observed with the camera. Figure~\ref{fig:usaf} (a) shows the output registered by the camera. For better visualization, the background was removed, and the image was binarized. A digital zoom of the central area can be seen in Figure~\ref{fig:usaf} (b), showing that the smallest blocks, corresponding to spacings of one and two pixels (lower left corner) can not be distinguished. Figures ~\ref{fig:usaf} (c) and (d) show integrated intensity profiles of the smallest distinguishable blocks, corresponding to the vertical lines with 3 and 4 pixels of spacing, respectively. Therefore, we expect our system to perform well down to a resolution of 3-4 pixels in full HD images.

\subsection{XOR operation efficiency}

\begin{figure}
    \centering
    \includegraphics[width=\linewidth]{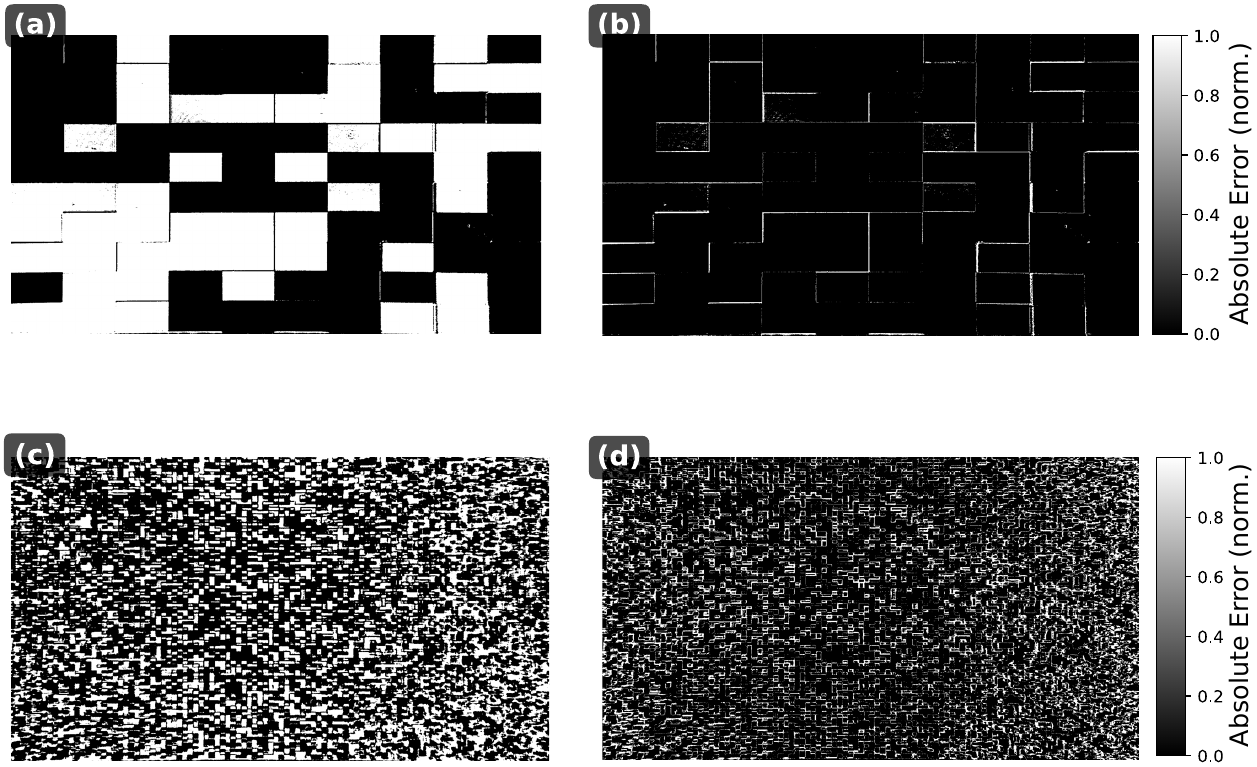}
    \caption{The experimentally obtained XOR operations results for $N\times N$ matrices encoded in the electric field wavefront, corresponding to (a) 100 and (c) 10000 elements. (b) and (d) show the absolute error between these images and the corresponding theoretical XOR operation.}
    \label{fig:xorimages}
\end{figure}

After evaluating its resolution, we assessed the system ability to perform the XOR operation on matrices of varying dimensions. For that, we displayed random binary matrices (composed of 0s and 1s) on SLM1, and a random binary matrix of the same dimension on SLM2. In a XOR-based encryption, the first image represents the information to be encrypted, while the second acts as a key. The recorded image is the result of the XOR operation between the two displayed matrices (see Equation~\ref{eq:Eout}). In order to be compared with theoretical expectations, the experimental images were normalized using the following formula:

\begin{equation}\label{normalization}
    I_{norm} = \frac{I_{exp} - I_{min}}{I_{max} - I_{min}},
\end{equation}

\noindent where $I_{min}$ denotes the image recorded with both SLMs set to 0-phase flat masks and $I_{max}$ corresponds to the image recorded with SLM1 at 0-phase and SLM-2 at $\pi$-phase flat masks. Since, locally, $I_{exp}$ is either equal to $I_{min}$ or to $I_{max}$, this normalization results in an image of only zeros and ones.

A set of typical normalized images is shown in Figures~\ref{fig:xorimages} (a) and (c) for different matrix dimensions. For the larger dimensions, the central area presents a more homogeneous representation of the matrix elements (rectangles) when compared to the borders of the image, indicating that the overlapping of the two original matrices is not ideal further from the center of the SLMs. This stems from slight misalignments and spherical aberration of the imaging lens. Besides, we note an edge enhancement for some adjacent matrix elements, that may be caused by diffraction at the strong refractive index step at the boundaries of contrasting rectangles in the original binary phase masks. All these aspects can be better visualized in Figures~\ref{fig:xorimages} (b) and (d), where the absolute value of the error between the normalized experimental image and the digitally computed result of the XOR operation is shown.

\begin{figure}
    \centering
    \includegraphics[width=.7\linewidth]{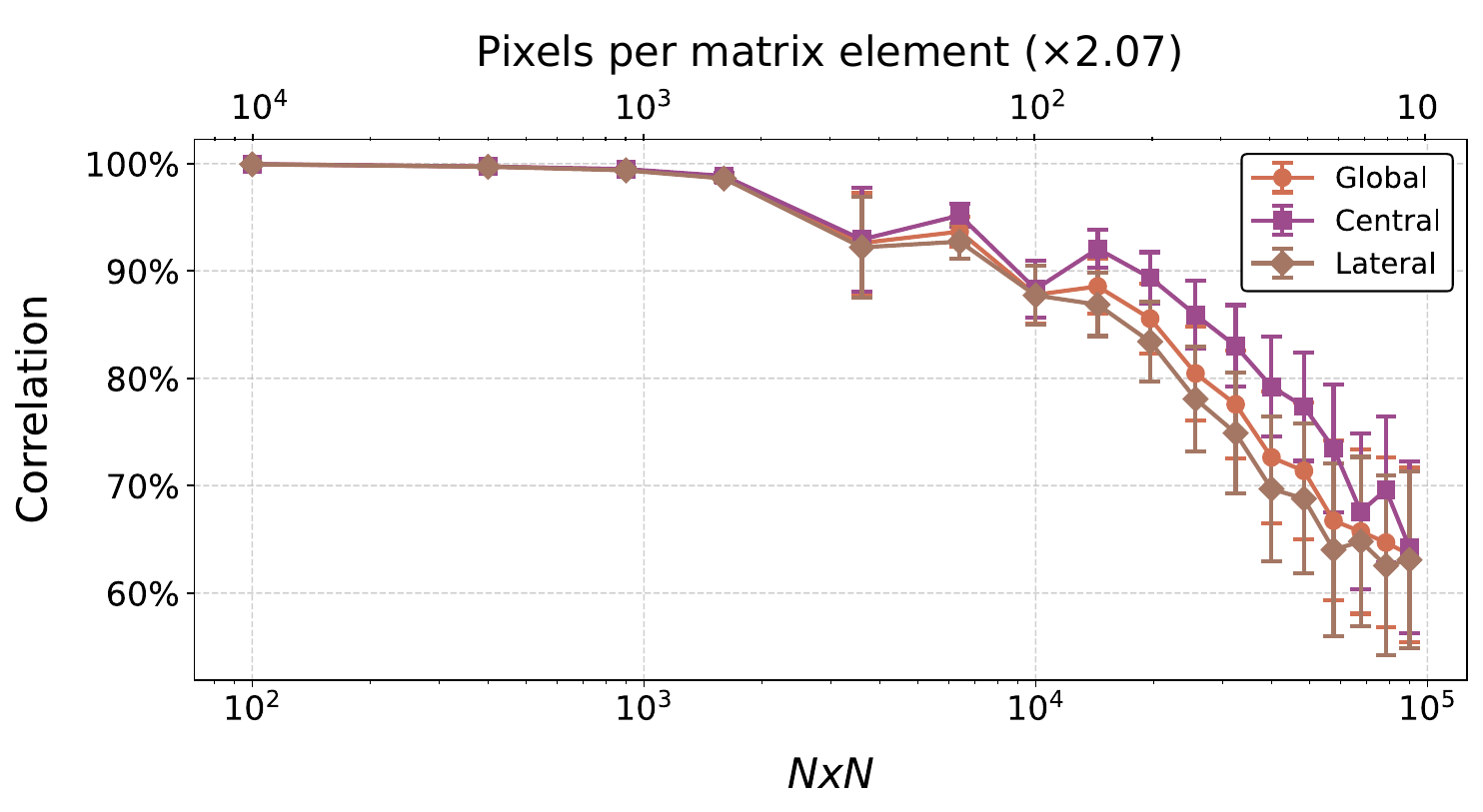}
    \caption{Correlation between experimental and theoretical XOR operations as a function of the total number of elements or bits (bottom axis) in the encoded matrices, and as a function of the number of pixels per bit (top axis).}
    \label{fig:correlation}
\end{figure}

We evaluated the accuracy of the optical XOR operation by computing the correlation between the normalized experimental images and the theoretical expectations, as shown in Figure \ref{fig:correlation}. For every matrix dimension, we used a sample size of 10 different images at each SLM, representing 10 independent trials. As theoretical expectations, we utilized the recorded images obtained when displaying the digitally computed result on SLM1 and a flat mask on SLM2, after normalization. In order to compare matrices elements and not image pixels, we have transformed the images in grayscale bitmaps by splitting them in element-sized rectangles, so that the bit gray level corresponds to the mean gray value of the pixels contained in the corresponding rectangle \cite{supplement}.

For the largest encoded matrix, 300 $\times$ 300 (corresponding to about 3-pixel high elements), the correlation was ($63\pm8$)\%. We note the different correlation levels obtained when analyzing the whole image (global), a centered squared region of 1080-pixels side (central) or the remaining parts closer to the borders (lateral).

The correlation quantifies the amount of bits for which the XOR operation is correctly applied in the optical setup, and thus maintain the qualitative behavior observed for the whole images (Figure~\ref{fig:xorimages}). On the one hand, Figure~\ref{fig:correlation} reveals that the bit error decreases for larger matrix dimensions, dropping below 20\% in the central region for matrices smaller than 200 $\times$ 200. On the other hand, this is due not only to technical limitations, but also to low information redundancy (top axis of Figure~\ref{fig:correlation}). The larger the matrix gets, the smaller the number of pixels used for representing each of its elements. Therefore, even tiny optical deleterious effects for image resolution (such as aberrations and misalignment) would increase the error rate.

\subsection{Image Encryption Application}

\begin{figure}
    \centering
    \includegraphics[width=\linewidth]{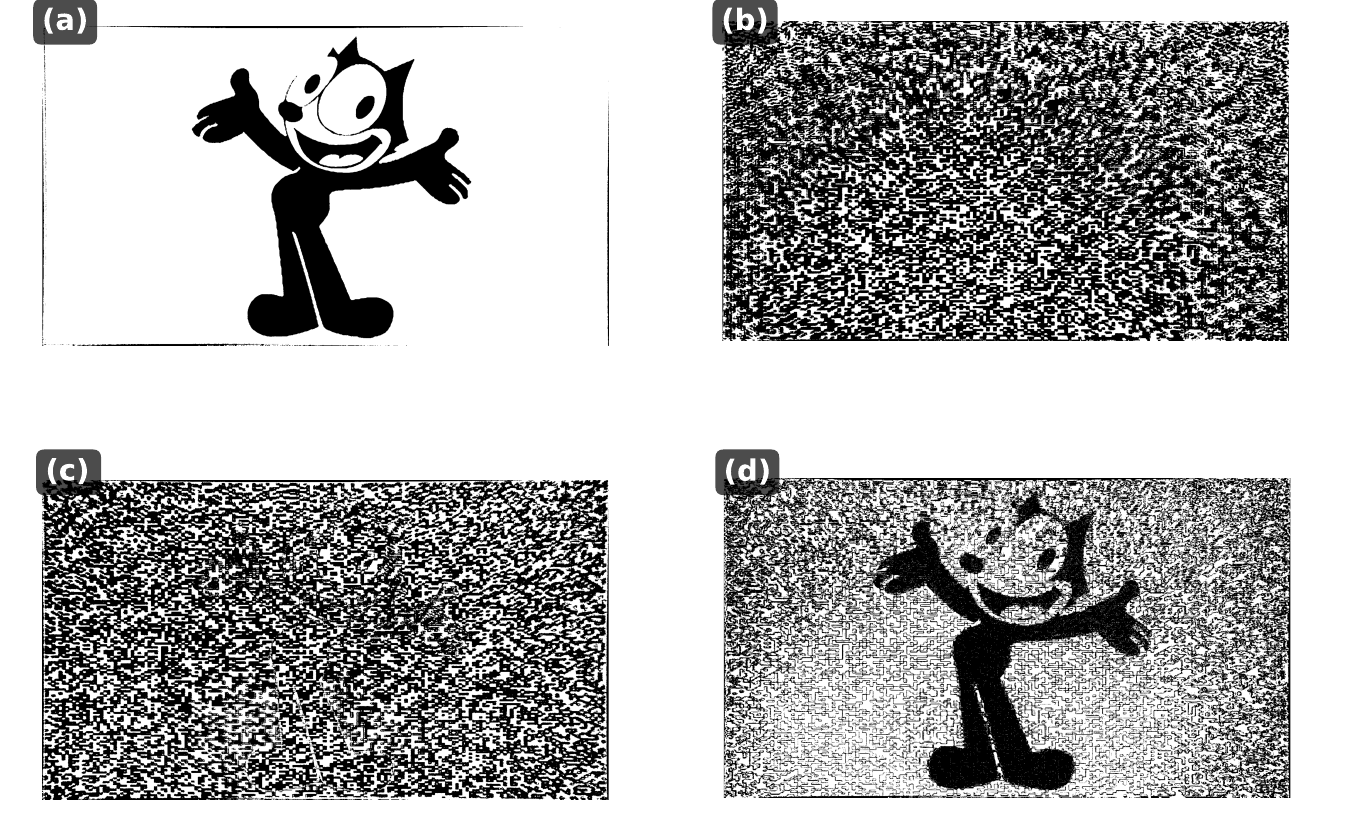}
    \caption{Performance of optical setup for image encryption and decryption with 164$\times$164 resolution. (a) The original image and (b) encryption key, as obtained experimentally. (c) The recorded, encrypted image after normalization and (d) the decrypted image obtained via computational XOR operation.}
    \label{fig:figure_panel}
\end{figure}

In order to assess the optical setup performance as an one-time-pad encryption system, we displayed a 164$\times$164 resolution black-and-white image on SLM1 and a random binary matrix as an encryption key on SLM2. Their recorded images, when displaying them on SLM1 while SLM had a flat mask, are shown in Figures~\ref{fig:figure_panel} (a) and (b), respectively. We applied the same normalization described above to the recorded, encrypted image, and the result is presented in Figure \ref{fig:figure_panel} (c). Then, we computationally decrypted the binarized image by utilizing the recorded image of the encryption key (Figure~\ref{fig:figure_panel} (b)). The decrypted image is shown in Figure~\ref{fig:figure_panel} (d). Figure \ref{fig:figure_panel} (c) demonstrates that the encryption scheme is able to successfully encipher binary information encoded in the light wavefront.

Furthermore, the correlation coefficient between the original and encrypted images was 6.17\%, indicating a high level of encryption security. After decryption, the correlation level increased to 42.55\%, suggesting partial reconstruction. Notably, the central region, which contains the main message (Felix the cat), exhibited a fairly high degree of correlation of 70.35\%, indicating better reconstruction fidelity in this area. In contrast, the lateral regions showed a significantly lower correlation of 22\%, suggesting a loss of information integrity in these parts of the image. This result is associated with the higher correlation values in the central region of the image when compared with its margins (Figure~\ref{fig:correlation}). This occurs because the SLM screen size is sufficiently large relative to the lens diameter, causing the peripheral portions of the light beam to illuminate areas of the lens where spherical aberration is more pronounced.

A last note on the encrypting capabilities of our optical setup concerns resolution. Both the image and the key used in this application have a resolution of 164$\times$164. The image to be encoded has mainly low-frequency information, which accounts for the degree of correlation of 70\%, above the level of about 50\% expected when inspecting Figure~\ref{fig:correlation}. Moreover, the results in Figure~\ref{fig:correlation} show that, should a twice-as-large matrix were to be used in this application, the correlation would drop by half. This is mainly due to errors introduced by pushing the system near its optical resolution, when both the aberrations at the borders and the lower information redundancy increase the errors.

\section{Conclusions}

In this work, we have analyzed the technical viability of an optical processing scheme capable of encoding and performing binary operations. Compared to existing optical encryption methods based on SLMs, our approach achieves a significantly higher bit capacity. To the best of our knowledge, this is the first demonstration of an image encoded using a key as large as 164$\times$164 in an optical XOR-based encryption scheme. In contrast, previous systems demonstrated only 21$\times$21 \cite{kumar2019image}, 30$\times$30 \cite{mogensen2000phase}, about 34$\times$34 \cite{tu2004optical} and 3-channel 11$\times$4 \cite{shikder2023image} encryption power.

We highlight that even when pushing our optical system near to its resolution limit of 3 pixels, about 64\% of the elements in the central portion of a large matrix (i.e., about 57600 bits) can be well encrypted. For ensuring a lower error rate, we find that information redundancy is relevant and poses a limit to the bit depth. In addition, low-frequency information (as the image used in the application we show here) can be encoded with about 30\% better accuracy than high-frequency ones, for being less sensitive to the imaging artifacts at the borders of matrix elements.

Despite its high encoding capacity, the system performance is sensitive to experimental parameters. Minor optical misalignments can degrade output quality, while larger lenses, though beneficial for spatial resolution, may still introduce distortions that compromise bit accuracy. Furthermore, environmental factors like dust and surface imperfections can generate unwanted diffraction, reducing image fidelity.

To address these limitations, future work will integrate aberration-compensated lenses, alongside advanced filtering techniques to suppress noise and diffraction artifacts. We will also study the possibility to implement and integrate error-correcting codes in our matrices. Such improvements can strengthen the system’s robustness, so that to position optical XOR encryption as a promising candidate for high-security optical data processing.

\section*{Acknowledgement}
This work has been supported by the following Brazilian research agencies: Conselho Nacional de Desenvolvimento Cient\'{\i}fico e Tecnol\'ogico (CNPq - DOI 501100003593), Coordena\c c\~{a}o de Aperfei\c coamento de Pessoal de N\'\i vel Superior (CAPES DOI 501100002322), Funda\c c\~{a}o de Amparo \`{a} Pesquisa do Estado de Santa Catarina (FAPESC - DOI 501100005667), Instituto Nacional de Ci\^encia e Tecnologia de Informa\c c\~ao Qu\^antica (INCT-IQ 465469/2014-0 and INCT-IQNano 406636/2022-2). AM is thankful to the French agency ANR (EVARISTE project ANR-24-CE23-1621) for financial support.

\vspace{1cm}
\textbf{Data availability}
All the data will be shared in a public repository after publication.

\textbf{Code availability}
All the codes for data analysis produced in association with our work will be shared in a public repository after publication.

\section*{Authors contributions}

M.G.D. implemented the optical processor, and conducted the measurements, with with contributions from R.M.A., P.H.S.R. and N.R.S. M.G.D. analyzed the data with contributions from A.M. and N.R.S. M.G.D. and N.R.S. wrote the manuscript with contributions from all authors. P.H.S.R. and N.R.S. conceived and directed the experimental setup and its evaluation. All authors discussed the results and the interpretation.


\bibliographystyle{ieeetr}
\bibliography{bib}

\end{document}


\maketitle

\section*{SI1: Experimental details}

Apart from the laser source and SLMs, the experimental setup was mounted using mostly Thorlabs products. In particular, we used a diode laser (Laserline iZi) delivering a linearly polarized, elliptical beam with diameter ratio of approximately 2.0, operating with a power of about 0.1 mW (5\% stability). The spatial light modulators (SLMs) are fabricated by Holoeye Photonics AG, with pixel pitches of 6.4~$\mu$m (LETO-3) and 3.74~$\mu$m (GAEA). The camera has a sCMOS monochromatic sensor (Thorlabs CS2100M-USB) an a pixel pitch of 5.04~$\mu$m. The imaging lenses were positioned between each of these three screens so that a proper magnification ensured the overlaying of pixels in a ratio of 1:4:1 relative to the LETO-3 display.

\begin{figure*}
\begin{center}
	\includegraphics[width=\textwidth]{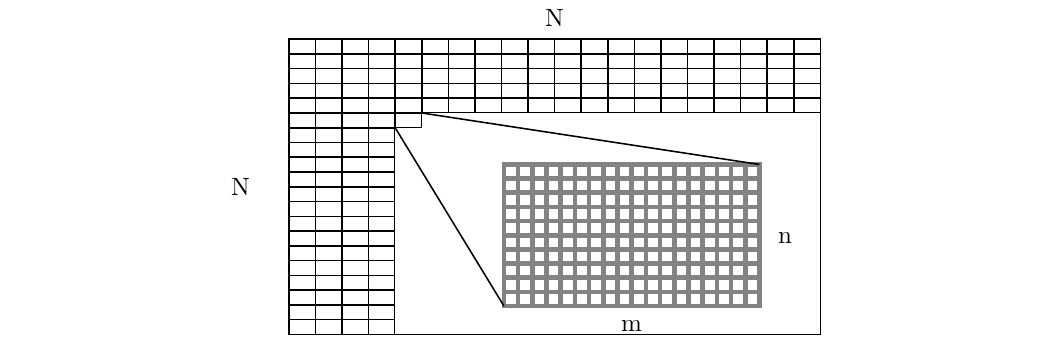}
	\caption{Schematic depiction of the splitting of a display in a $N\times N$ matrix. The inset is a zoomed element of the matrix, composed by $m\times n$ pixels in a Full HD resolution image.}
	\label{supplfig:matrixDisplay}
\end{center}
\end{figure*}

The binary matrices used for testing the optical processor were generated by filling a bitmap split in $N\times N$ blocks with either 0 or 1 (randomly chosen). The bitmap grid had a fixed size of $1920\times1080$ pixels (Full-HD resolution), so that each block corresponding to a matrix element contained $m=1920/N$ times $n=1080/N$ pixels (Fig.~\ref{supplfig:matrixDisplay}).

The preparation of random binary matrices, as well as the controlling of both SLMs and camera acquisition were automatized using a self-developed Python code.


\section*{SI2: Image acquisition and normalization}

For recording images of optical matrices, we fixed the exposure time in each experimental series. A series consisted of a sequence of matrices (operations or not) of a given dimension, along with two images recorded when both SLMs displayed the same blank images (corresponding to phases of either 0 or $\pi$ for each of them) or opposite blank images (corresponding one to phase 0, and the second to phase $\pi$).

Within a series, the recorded images $I_{exp}$ of matrices were all normalized according to the following procedure. The image obtained using the same blank images, $I_{min}$, and the one corresponding to opposite blank images, $I_{max}$, were considered as a map of the minimum and maximum values per pixel. The normalized images were thus calculated by

\begin{equation}
I_{norm} = \dfrac{I_{exp}-I_{min}}{I_{max}-I_{min}}.
\label{eq:normImg}
\end{equation}

Two examples of recorded and corresponding normalized images are shown at the bottom rows of Fig.~\ref{supplfig:imagesXOR} in the following Section. The normalization was handled by a Python code \footnote{All data and image-processing codes will be shared in a public repository after publication.}.


\section*{SI3: Result of XOR operation -- theoretical \textit{versus} recorded image of it}

The optical processor is able to perform the XOR operation under experimental limitations, like optical misalignment, imaging aberrations and display pixel pitch. In order to make a fair comparison between the operations fully realized by the processor and theoretical expectations, we have to take into account such limitations. A practical way to achieve that is to display the mathematically obtained result of a XOR operation in one of the SLMs and record an image of it, so that the experimental setup may affect this result in a similar manner as it introduces artifacts in the optically computed outcome.

\begin{figure*}[h!]
\begin{center}
	\includegraphics[width=\textwidth]{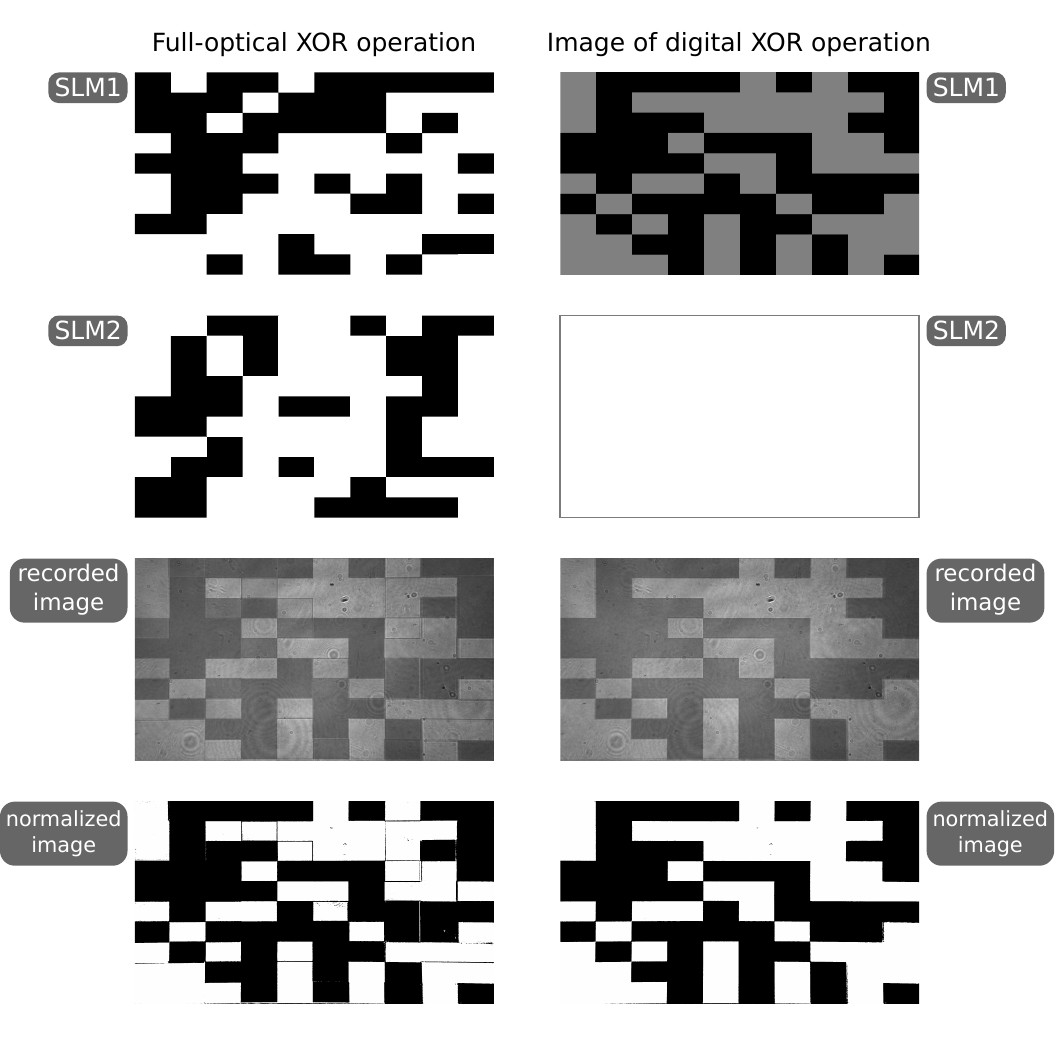}
	\caption{Examples of XOR operation results obtained in the optical setup. Each vertical panel shows, from top to bottom, the matrices displayed in a single SLM, the images as recorded by the camera and after normalization. Left panel: XOR operation realized by the optical processor. Right panel: theoretically calculated XOR operation result, when recorded by the same optical setup.}
	\label{supplfig:imagesXOR}
\end{center}
\end{figure*}

Figure~\ref{supplfig:imagesXOR} shows the images of XOR operation results between a pair of matrices realized directly in the optical processor (left panel) and digitally (right panel). The aforementioned experimental artifacts are visible as ring-like patterns and diffraction fringes between some matrix elements in both recorded images (third row). They are mostly removed by the normalization procedure (fourth row), and the images of XOR-operated matrices can be compared.

\section*{SI4: Correlation as a measure of XOR operation efficiency}

We have shown that the correlation between optically XOR-operated matrices and the expected theoretical result remained above 60\% up to $300\times300$ matrices (Fig.~4 in the main text). Here, we give more details on how we obtain such measures.

Firstly, we take all experimental images and reduce them to bitmaps of the same size of the operated matrices. For that, we take each block of $m\times n$ pixels (see Fig.~\ref{supplfig:matrixDisplay}) and replace it by a single bit whose gray value corresponds to the mean intensity of the block. For instance, the images at the bottom of Fig.~\ref{supplfig:imagesXOR} were transformed into $10\times10$ grayscale bitmaps.

The correlation between the so-obtained bitmaps of the results of XOR operations performed by the optical processor and obtained mathematically was then computed by:

\begin{equation}
\text{corr}(x,y) = \dfrac{\sum_{i=1}^{n} (x_i - \bar{x}) (y_i - \bar{y})}{\sqrt{\sum_{i=1}^{n} (x_i - \bar{x})^2}\ \sqrt{\sum_{i=1}^{n} (y_i - \bar{y})^2}},
\label{eq:correlation}
\end{equation}

\noindent where $\bar{x}$ and $\bar{y}$ are the mean values of $x$ and $y$. Here, the variables $x$ and $y$ are the bitmaps of the results of the XOR operation realized experimentally and digitally, respectively. Such a correlation is usual for quantitatively assessing the similarity between digital images.

The image conversion to a grayscale bitmap was done using Java, while the correlation was computed using Python \footnote{All data and image-processing codes will be shared in a public repository after publication.}.